\documentclass{emulateapj}
\usepackage{multirow}
\usepackage{color}
\usepackage[normalem]{ulem}
\usepackage{amssymb}

\begin{document}

\title{FAR-ULTRAVIOLET SPECTRAL IMAGES OF THE VELA SUPERNOVA REMNANT: SUPPLEMENTS AND COMPARISONS WITH OTHER WAVELENGTH IMAGES}

\author{Il-Joong Kim\altaffilmark{1}}
\author{Kwang-Il Seon\altaffilmark{1}}
\author{Kyoung-Wook Min\altaffilmark{2}}
\author{Wonyong Han\altaffilmark{1}}
\author{Jerry Edelstein\altaffilmark{3}}

\altaffiltext{1}{Korea Astronomy and Space Science Institute, Daejeon 305-348, Republic of Korea; ijkim@kasi.re.kr}
\altaffiltext{2}{Korea Advanced Institute of Science and Technology, Daejeon 305-701, Republic of Korea}
\altaffiltext{3}{Space Sciences Laboratory, University of California, Berkeley, CA 94702, USA}

\begin{abstract}
We present the improved far-ultraviolet (FUV) emission-line images of the entire \object{Vela} supernova remnant (SNR) using newly processed SPEAR/FIMS data. The incomplete \ion{C}{3} $\lambda$977 and \ion{O}{6} $\lambda\lambda$1032, 1038 images presented in the previous study are updated to cover the whole region. The \ion{C}{4} $\lambda\lambda$1548, 1551 image with a higher resolution and new images at \ion{Si}{4} $\lambda\lambda$1394, 1403, \ion{O}{4}{]} $\lambda$1404, \ion{He}{2} $\lambda$1640.5, and \ion{O}{3}{]} $\lambda\lambda$1661, 1666 are also shown. Comparison of emission line ratios for two enhanced FUV regions reveals that the FUV emissions of the east enhanced FUV region may be affected by nonradiative shocks of another very young SNR, the Vela Jr. SNR (RX J0852.0--4622, G266.6-1.2). This result is the first FUV detection that is likely associated with the Vela Jr. SNR, supporting previous arguments that the Vela Jr. SNR is close to us. The comparison of the improved FUV images with soft X-ray images shows that a FUV filamentary feature forms the boundary of the northeast-southwest asymmetrical sections of the X-ray shell. The southwest FUV features are characterized as the region where the \object{Vela} SNR is interacting with slightly denser ambient medium within the dim X-ray southwest section. From a comparison with the H$\alpha$ image, we identify a ring-like H$\alpha$ feature overlapped with an extended hot X-ray feature of similar size and two local peaks of \ion{C}{4} emission. Their morphologies are expected when the H$\alpha$ ring is in direct contact with the near or far side of the \object{Vela} SNR.
\end{abstract}


\section{INTRODUCTION}

The \object{Vela} supernova remnant (SNR) is one of the brightest Galactic SNRs because of its proximity ($\sim$290 pc; Caraveo et al. 2001; Dodson et al. 2003) and large angular diameter ($\sim$8$\arcdeg$; Aschenbach et al. 1995). It is also a middle-aged SNR ($\sim$10$^{4}$ yr; Reichley et al. 1970), which is expected to brighten in various wavelength domains. It has been extensively observed and analyzed in almost all wavelength domains. The {\it ROSAT} X-ray image \citep{Aschenbach95} revealed the previously unknown faint western part, and also discovered six extended X-ray features outside the blast-wave front (Shrapnels A--F). From spatially resolved X-ray spectral analysis, \citet{Lu00} presented global maps of emission measure, temperature, and absorption column density ($N_{\mathrm H}$). The absorption column density ranges from 5.0$\times$10$^{19}$ to 6.0$\times$10$^{20}$ cm$^{-2}$. They also found that most \ion{H}{1} is located behind the \object{Vela} remnant, particularly in the bright northeastern part, and therefore the X-ray brightness of the \object{Vela} SNR is not significantly affected by $N_{\mathrm H}$ variations. \citet{Dubner98} studied the distribution and kinematics of the \ion{H}{1} in the \object{Vela} SNR's direction and found the presence of a $\sim$7$\arcdeg$ circular \ion{H}{1} shell surrounding the remnant. They suggested that the \object{Vela} SNR is expanding in the pre-existing \ion{H}{1} bubble which is partially swept-up. The CO \citep{Moriguchi01} and dust \citep{Nichols04} studies also indicate that the \object{Vela} SNR is actively interacting with a cloudy and dusty ambient medium. Recently, \citet{Sushch11} developed a hydrodynamical model for the interaction of the \object{Vela} SNR and the $\gamma^{2}$ Velorum stellar wind bubble (SWB). They proposed that the \object{Vela} SNR exploded near the boundary of the $\gamma^{2}$ Velorum SWB and the observed northeast-southwest asymmetry of the \object{Vela} SNR may be due to abrupt changes in the physical parameters of the ISM at the boundary (see Figure 2(a)).

The proximity and negligible foreground material (low $N_{\mathrm H}$) enable the detection of far-ultraviolet (FUV) emission from the \object{Vela} SNR. Various FUV emission lines have been detected in the central, north edge, and Shrapnel D regions \citep{Blair95,Raymond81,Raymond97,Sankrit01,Sankrit03}. These emission lines were interpreted as the existence of $\sim$100--180 km s$^{-1}$ shocks driven by a blast-wave. However, the FUV observations were restricted to within only a small portion of the remnant, such as individual shock fronts. The first global FUV observation of the \object{Vela} SNR was carried out by the Spectroscopy of Plasma Evolution from Astrophysical Radiation (SPEAR), also known as Far-Ultraviolet Imaging Spectrograph (FIMS). The early results of the SPEAR/FIMS data analysis were presented in \citet{Nishikida06}. \citet{Nishikida06} showed the complete-coverage \ion{C}{4} $\lambda\lambda$1548, 1551 and \ion{Si}{4}+\ion{O}{4}{]} $\lambda\lambda$1400, 1403 (unresolved) images together with the incomplete-coverage \ion{C}{3} $\lambda$977 and \ion{O}{6} $\lambda\lambda$1032, 1038 images. These FUV images extend $\sim$8$\arcdeg$, which is similar to the X-ray image. The global FUV spectra were found to be generally consistent with previous FUV observations. It was also noted that the global FUV data indicate inhomogeneous shock-induced emission from the \object{Vela} SNR surface.

In the present study, we update the global FUV spectral images by using the newly processed SPEAR/FIMS data. In addition to the emission-line images presented in \citet{Nishikida06}, the \ion{He}{2} $\lambda$1640.5 and \ion{O}{3}{]} $\lambda\lambda$1661, 1666 images have been newly obtained. The images cover the whole region of the \object{Vela} SNR and reveal more detailed features than those in \citet{Nishikida06}. We compare the newly produced FUV images with the X-ray and H$\alpha$ images, and examine how the \object{Vela} SNR evolves and interacts with the ambient medium on a global scale. We also present the FUV emission-line luminosities estimated for a few subregions as well as the whole region, which are compared with the X-ray luminosities and those of the \object{Cygnus Loop}. Additionally, we examine the FUV line ratio diagrams to investigate the contribution of the Vela Jr. SNR (RX J0852.0--4622, G266.6-1.2), known as an younger SNR overlapped with the \object{Vela} SNR in projection \citep{Aschenbach98}.

\section{DATA REDUCTION}

SPEAR/FIMS is the primary payload on the first Korean Science and Technology Satellite, {\it STSAT-1}, launched on 2003 September 27. To observe large-scale diffuse FUV emission lines from the interstellar medium (ISM), SPEAR/FIMS was designed as a dual-channel FUV imaging spectrographs: the short wavelength channel (S-channel; 900--1150 \AA{}) with 4$\arcdeg$.0$\times$4$\arcmin$.6 field of view and long wavelength channel (L-channel; 1340--1750 \AA{}) with 7$\arcdeg$.4$\times$4$\arcmin$.3 field of view. Its spectral and angular resolutions are $\lambda/\Delta\lambda\sim550$ and 5$\arcmin$, respectively. The SPEAR/FIMS mission, the instruments, its on-orbit performance, and the basic processing of the data are described in detail by Edelstein et al. (2006a, 2006b).

We used the data from a total of 115 orbits including 16 orbits used in \citet{Nishikida06}. Of the 115 total orbits, 77 orbits observed in the sky survey observational mode were newly included so as to complete the region uncovered for the S-channel in \citet{Nishikida06}. Instead of rejecting all data with poor attitude knowledge ($>$30$\arcmin$) as done in \citet{Nishikida06}, we increased the attitude accuracy up to 5$\arcmin$ by using automated software correction similar to that used in \citet{Seon06}. The positions of bright stars were corrected to match their positions listed in the {\it TD-1} catalog \citep{Thompson78}. Then, the position errors of all photons were corrected by linearly interpolating from those of the bright stars. We removed the data recorded when the count rate was high ($>$1000 counts s$^{-1}$) because these events are mostly associated with bright stars. The photons of bright stars are scattered off by interstellar dust and spread over the whole field of view, dominating over any diffuse emission \citep{Seon11}. For the L-channel data, only the 1360--1690 \AA{} portion was selected so as to exclude the strong geocoronal airglow line \ion{O}{1} $\lambda$1356 and long wavelength part (1690--1750 \AA{}) with a relatively high detector background. Eventually, the processed data included a total of 4.4$\times$10$^{6}$ events in the L-channel and 2.0$\times$10$^{5}$ events in the S-channel. To obtain images and spectra, we adopted the HEALPix scheme \citep{Gorski05} with a resolution parameter \texttt{Nside} = 2048, corresponding to a pixel size of $\sim$1$\arcmin$.72, and 1 \AA{} wavelength bins for the L-channel and \texttt{Nside} = 1024 and 0.5 \AA{} for the S-channel. In each channel, the effects of bright stars were further reduced by masking pixels whose continuum intensities are 10 times higher than the median value of the whole region: $\sim$2\% of the total pixels.

\section{DATA ANALYSIS AND RESULTS}

To produce each emission-line image, the spectral region around each emission line was taken and each portion was fitted with a line profile (or line profiles) of Gaussian shape, plus, a linear background continuum for each pixel. The centers and widths of the Gaussian functions were fixed using the calibrated line centers and the SPEAR/FIMS spectral resolutions, respectively. The \ion{O}{6} $\lambda\lambda$1031.9, 1037.6 and \ion{C}{4} $\lambda\lambda$1548.2, 1550.8 doublet lines have an intrinsic line ratio of 2:1, but resonant scattering makes the ratio approach to 1:1 in an optically thick case \citep{Long92}. According to this, we set a line-ratio parameter to vary between 2:1 and 1:1. The \ion{O}{3}{]} lines were assumed to be formed in a doublet at 1660.8 and 1666.1 \AA{} with a 0.34:1 ratio of their statistical weights. Although \citet{Nishikida06} did not deblend the \ion{Si}{4} + \ion{O}{4}{]} lines, we attempted to fit these line complexes by assuming the following. The \ion{O}{4}{]} lines were assumed to be a quintet of lines (at 1397.2, 1399.8, 1401.2, 1404.8, and 1407.4 \AA{}). The \ion{Si}{4} $\lambda\lambda$1393.8, 1402.8 doublet lines have line ratios as the \ion{O}{6} and \ion{C}{4} doublet lines do. Since the weaker 1402.8 \AA{} line could not be resolved from the \ion{O}{4}{]} quintet lines, only the stronger 1393.8 \AA{} line was fitted together with the \ion{O}{4}{]} quintet lines. Then, the weaker line strength was assumed to be half the strength of the stronger line, although this is valid only for an optically thin case. This amount was subtracted from the fitting result of the \ion{O}{4}{]} quintet lines and added to that of the \ion{Si}{4} $\lambda$1393.8 line. This has not largely affected the fitting results (particularly for the \ion{O}{4}{]} line values) because the \ion{Si}{4} $\lambda$1393.8 line strength was mostly much lower than the integrated strength of the \ion{O}{4}{]} quintet lines (see Table 1). Before fitting, the pixel size was increased up to \texttt{Nside} = 256 (corresponding to a pixel size of $\sim$13$\arcmin$.7) for better statistics; the brightest emission-line image, \ion{C}{4}, could be produced with a smaller pixel size of \texttt{Nside} = 512. The resulting images had some pixels with very high intensity relative to their adjacent pixels, which are mostly related to unremoved point sources. These pixels contain only a small number of photon events and have been further excluded in the final images. The solid lines in Figure 1 denote the outlines of the excluded pixels.

\begin{figure}[t]
\begin{centering}
\includegraphics[scale=0.5]{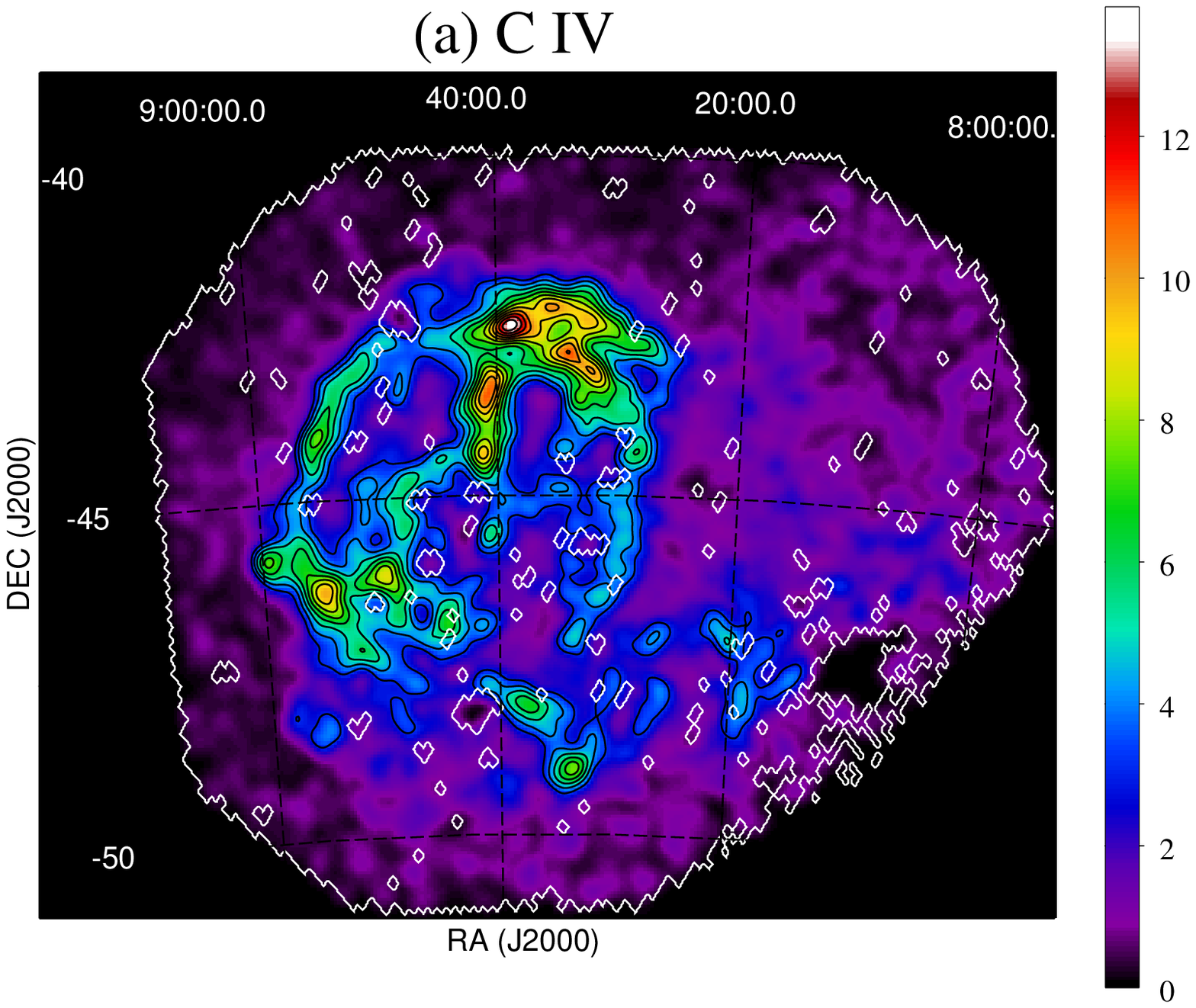}
\par\end{centering}
\begin{centering}
\includegraphics[scale=0.25]{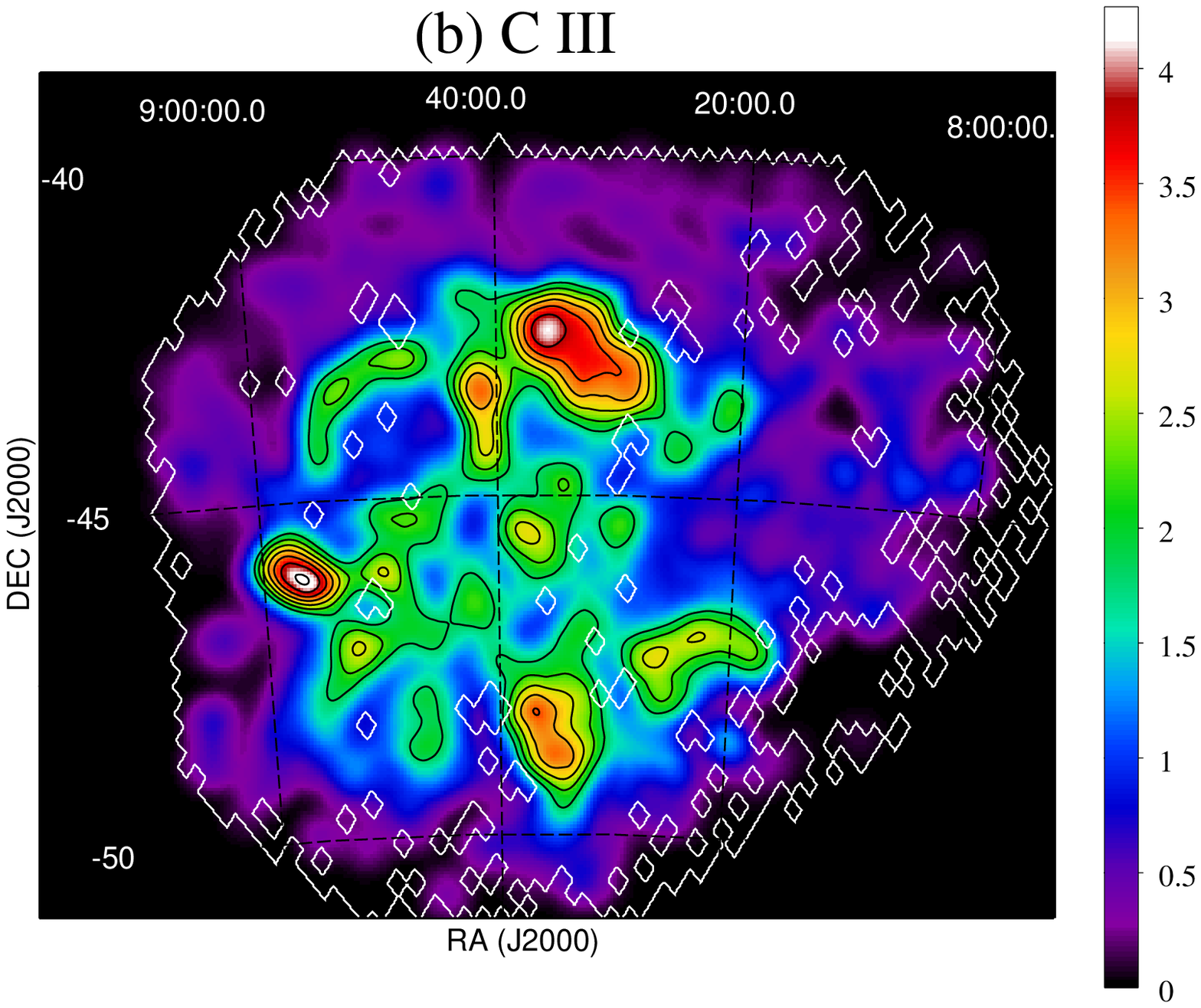}\quad\includegraphics[scale=0.25]{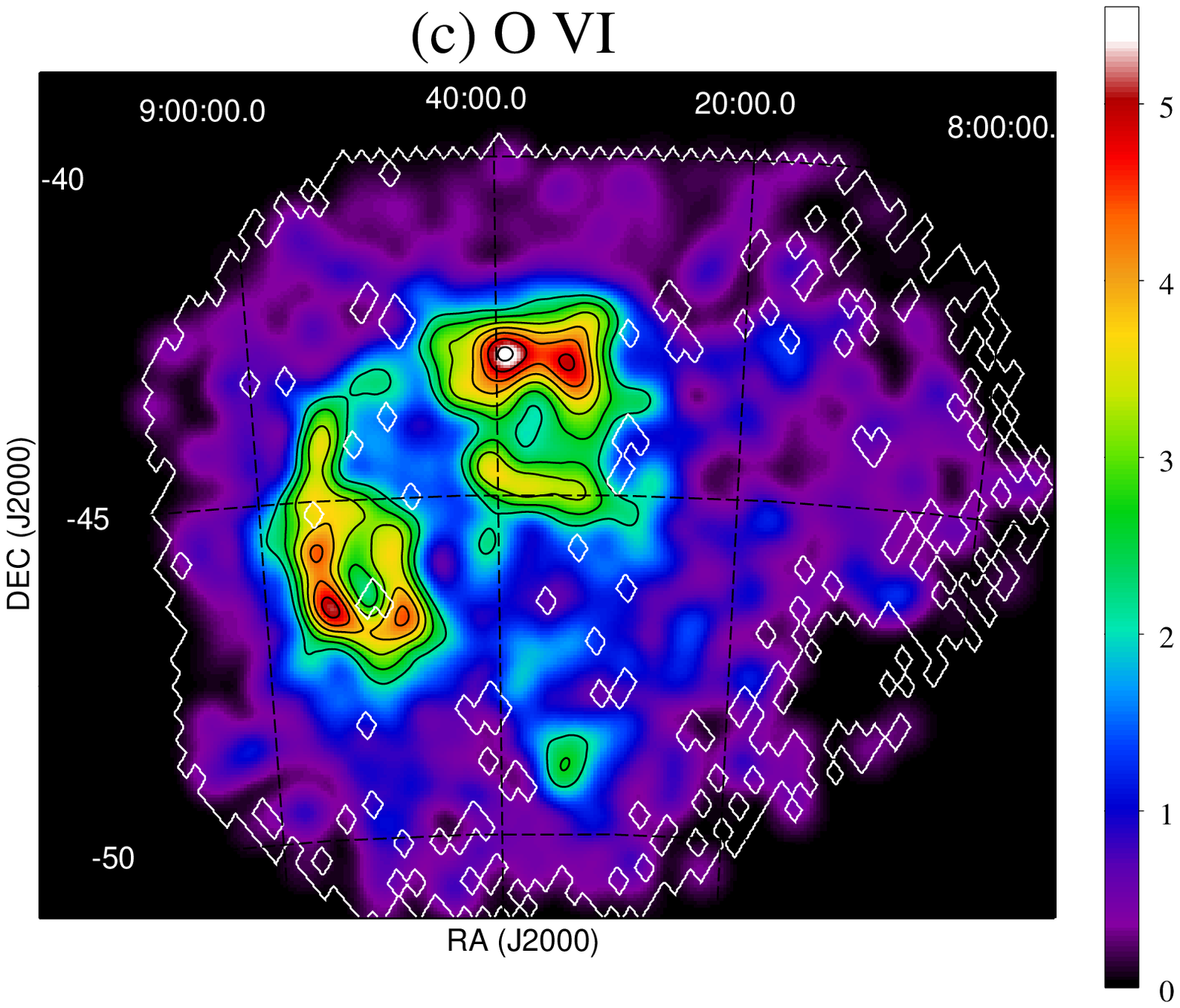}
\par\end{centering}
\begin{centering}
\includegraphics[scale=0.25]{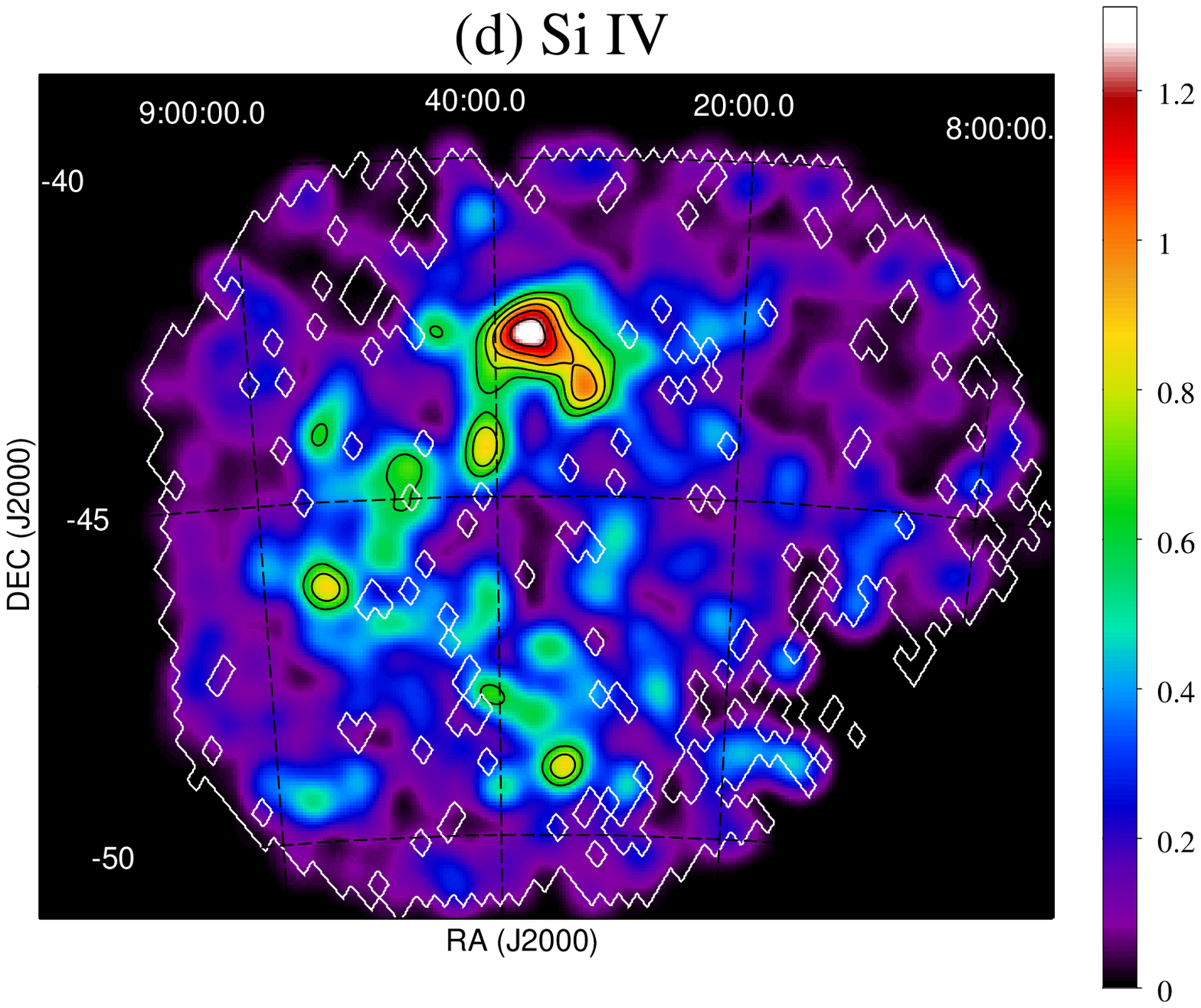}\quad\includegraphics[scale=0.25]{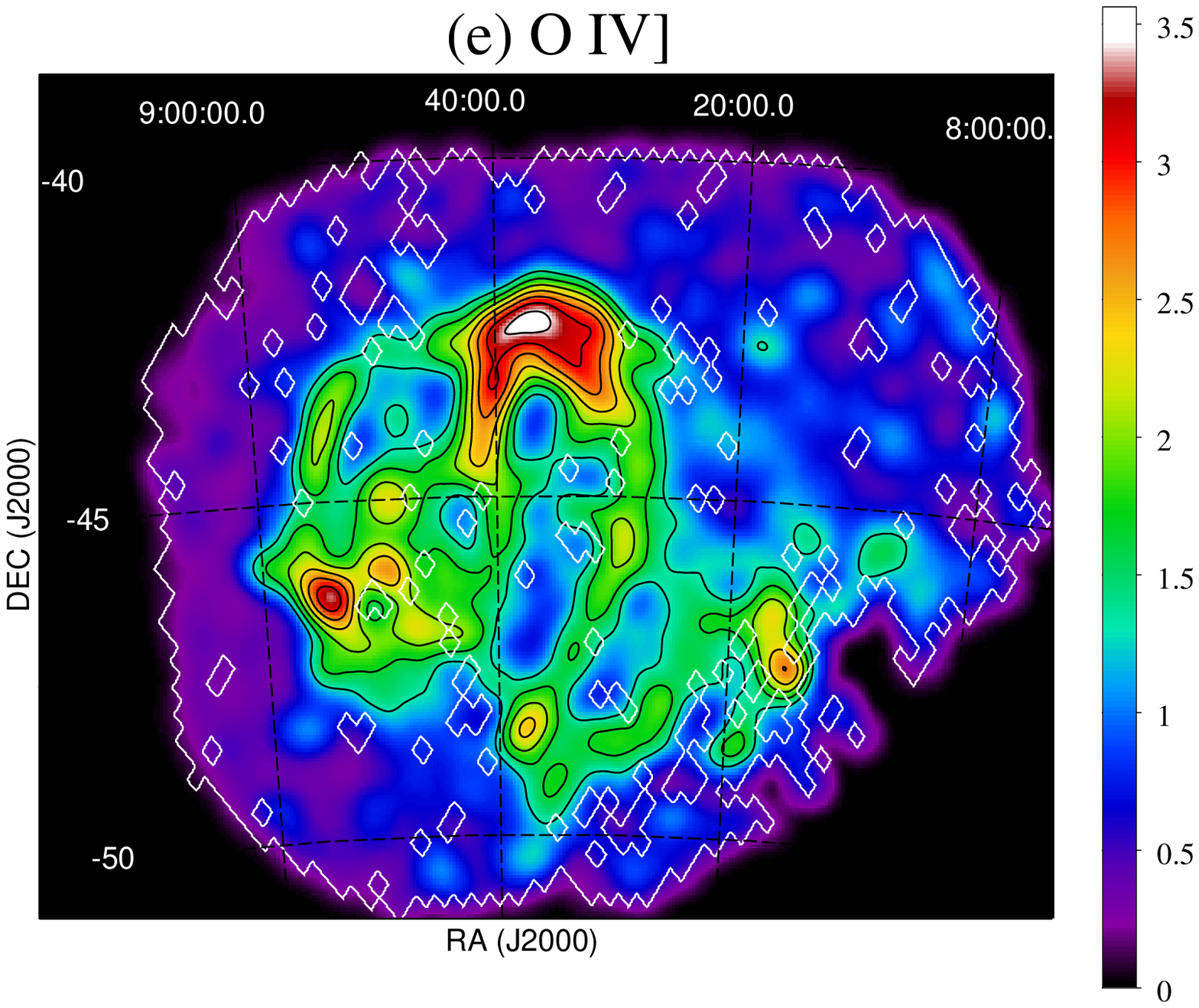}
\par\end{centering}
\begin{centering}
\includegraphics[scale=0.25]{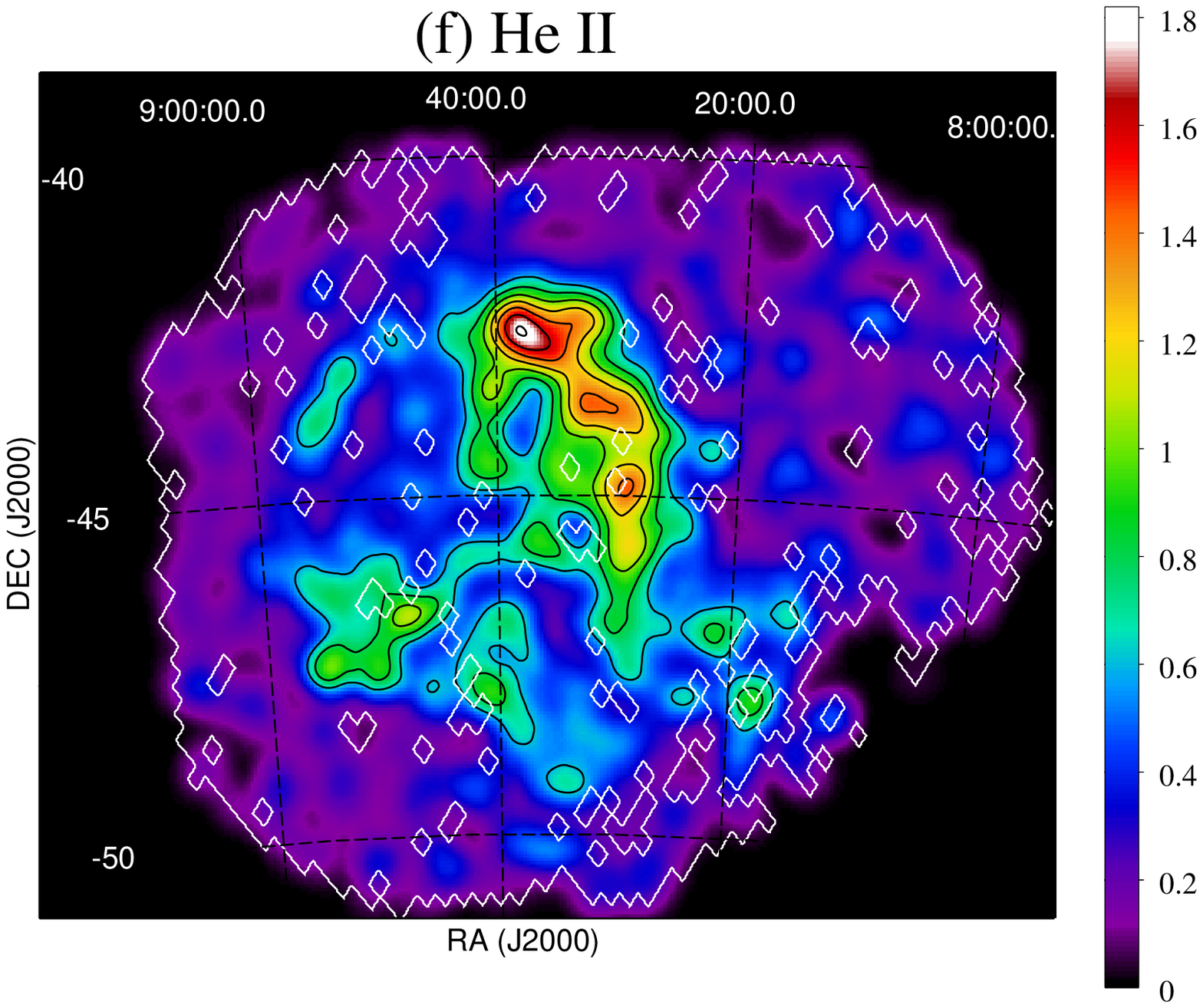}\quad\includegraphics[scale=0.25]{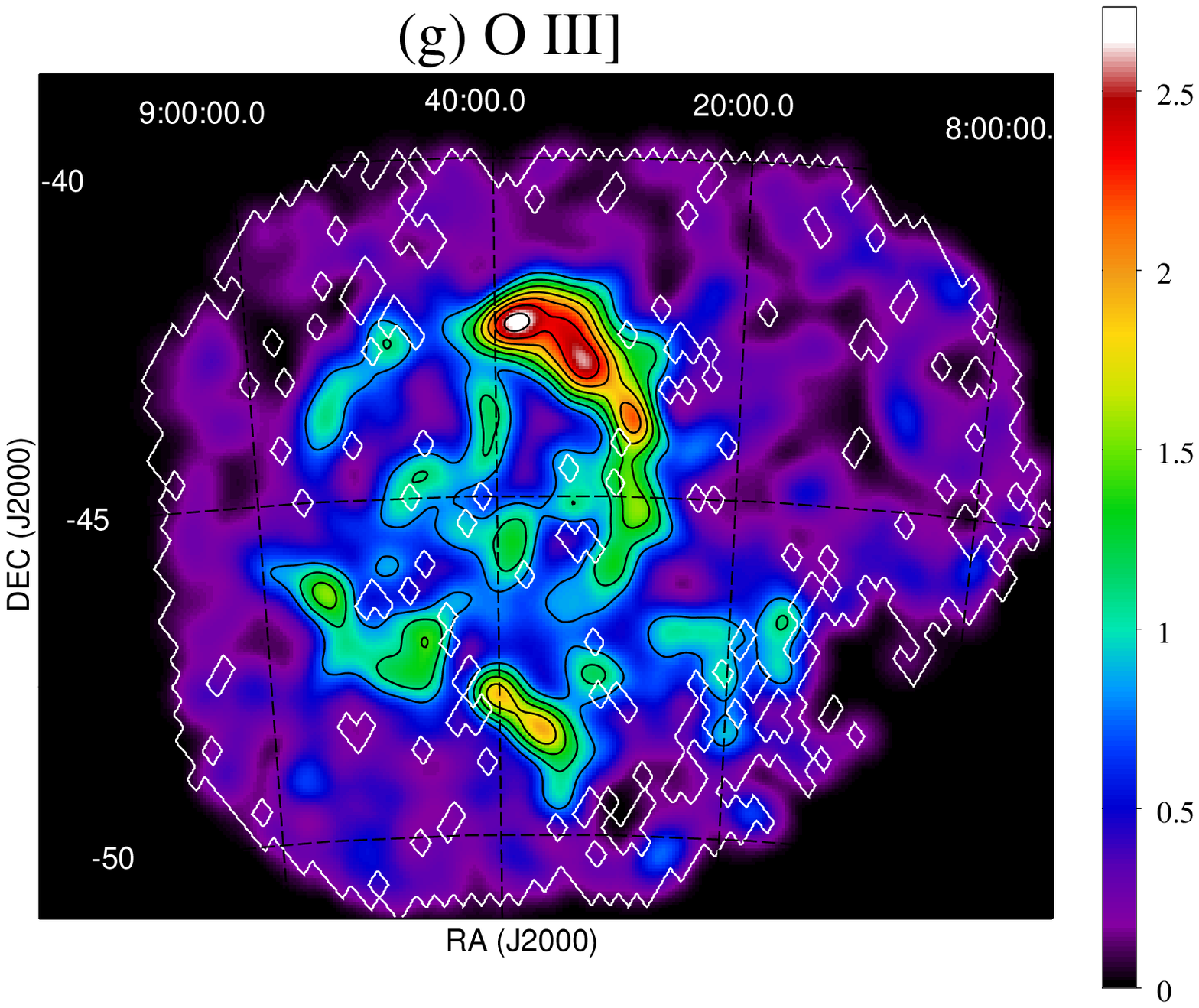}
\par\end{centering}
\textbf{Figure 1.} SPEAR/FIMS (a) \ion{C}{4} $\lambda\lambda$1548, 1551, (b) \ion{C}{3} $\lambda$977, (c) \ion{O}{6} $\lambda\lambda$1032, 1038, (d) \ion{Si}{4} $\lambda\lambda$1394, 1403, (e) \ion{O}{4}{]} $\lambda$1404, (f) \ion{He}{2} $\lambda$1640.5, and (g) \ion{O}{3}{]} $\lambda\lambda$1661, 1666 emission-line images of the \object{Vela} SNR. The units of the color bars are 10$^{-6}$ erg s$^{-1}$ cm$^{-2}$ sr$^{-1}$. Contour levels in the same units are stepped by seven (or eleven for (a)) equal intervals and range (a) from 3.0 to 13.0, (b) from 1.8 to 4.2, (c) from 2.1 to 5.4, (d) from 0.6 to 1.5, (e) from 1.3 to 3.4, (f) from 0.6 to 1.8, and (g) from 0.8 to 2.6. The values are not corrected for interstellar extinction. The solid lines (colored white in the online version) indicate the outlines of the masked pixels.
\end{figure}

Figure 1 gives the final results of the seven emission-line images. They have been smoothed using a Gaussian function with a kernel radius of 30$\arcmin$ (or 15$\arcmin$ for \ion{C}{4}). In each image, seven (or eleven for \ion{C}{4}) equally-spaced contour levels were overlaid. The pixels enclosed by the lowest-level contours have a signal-to-noise ratio (S/N) of $>$3 (except that S/N $>$2 for the \ion{C}{3} and \ion{O}{6} lines belonging to the S-channel with relatively low sensitivity). Based on Figure 1(a) for the \ion{C}{4} image with the highest spatial resolution, several noticeable features are identified. There are two enhanced regions: the north (around $\alpha \sim 8^{\mathrm h}35^{\mathrm m}, \delta \sim -42\arcdeg30\arcmin$) and the east (around $\alpha \sim 8^{\mathrm h}52^{\mathrm m}, \delta \sim -46\arcdeg20\arcmin$) enhanced FUV regions. Both regions appear in all the other emission-line images, of which the \ion{O}{6} image shows them most outstandingly. The north enhanced FUV region is the most prominent region and has already been mentioned in \citet{Nishikida06}. However, the improved \ion{C}{4} image in Figure 1(a) reveals two concentric arc-like features that could not be resolved in \citet{Nishikida06}. In addition to the enhanced FUV regions, we note three filamentary features running north-south: the northeast limb filament from $\alpha \sim 8^{\mathrm h}48^{\mathrm m}, \delta \sim -42\arcdeg40\arcmin$ to $\alpha \sim 8^{\mathrm h}56^{\mathrm m}, \delta \sim -44\arcdeg40\arcmin$ (FUV filament A), the east central filament from $\alpha \sim 8^{\mathrm h}40^{\mathrm m}, \delta \sim -43\arcdeg$ to $\alpha \sim 8^{\mathrm h}42^{\mathrm m}, \delta \sim -44\arcdeg40\arcmin$ (FUV filament B), and the west central filament from $\alpha \sim 8^{\mathrm h}28^{\mathrm m}, \delta \sim -44\arcdeg$ to $\alpha \sim 8^{\mathrm h}34^{\mathrm m}, \delta \sim -47\arcdeg15\arcmin$ (FUV filament C). FUV filaments A and B are shown in all emission-line images, and also appear in the figures of \citet{Nishikida06}. FUV filament C, however, is somewhat faint and has been newly identified in the present study. The newly obtained \ion{O}{4}{]}, \ion{He}{2}, and \ion{O}{3}{]} images confirm this filament more clearly, as can be seen in Figures 1(e), (f), and (g). Finally, there are southwest FUV features with a peak intensity at $\alpha \sim 8^{\mathrm h}34^{\mathrm m}, \delta \sim -49\arcdeg$. As shown in Figure 1(a), they extend towards the west beyond FUV filament C. While the peak position is seen in almost all the emission-line images, the west extended part is not clear in some images, particularly in the \ion{O}{6} image.

\begin{figure}[t]
\begin{centering}
\includegraphics[scale=0.25]{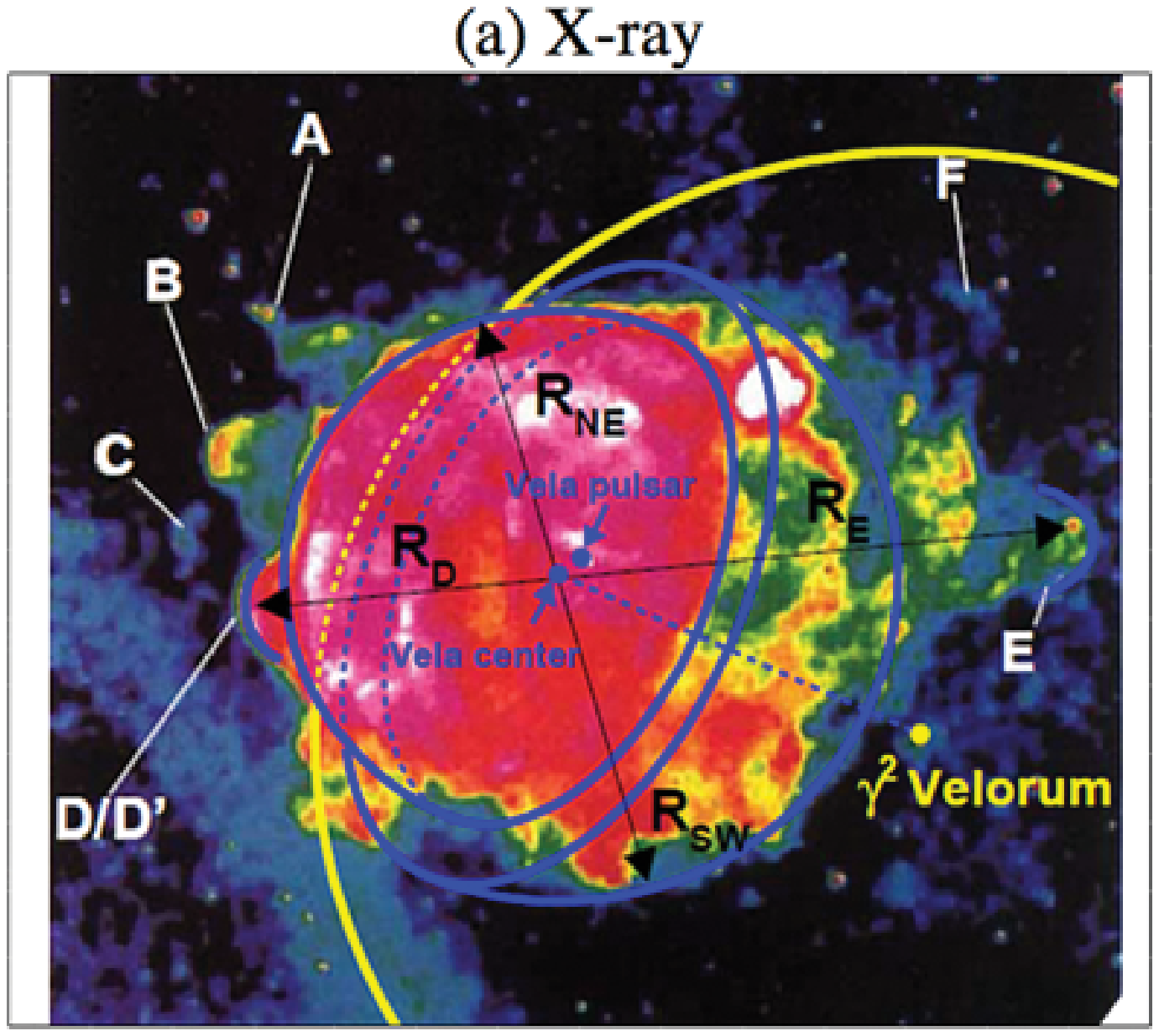}\quad\includegraphics[scale=0.25]{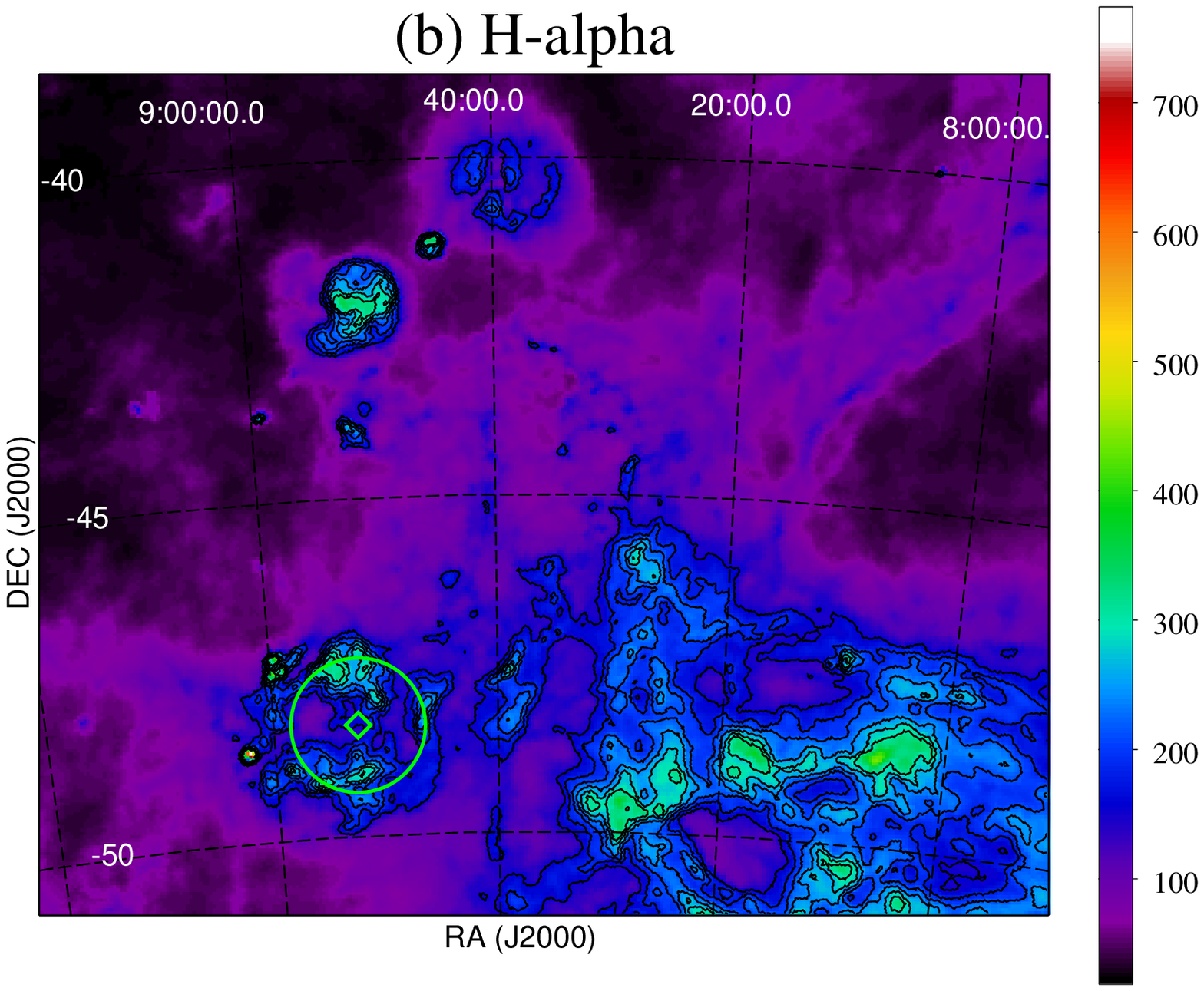}
\par\end{centering}
\begin{centering}
\includegraphics[scale=0.25]{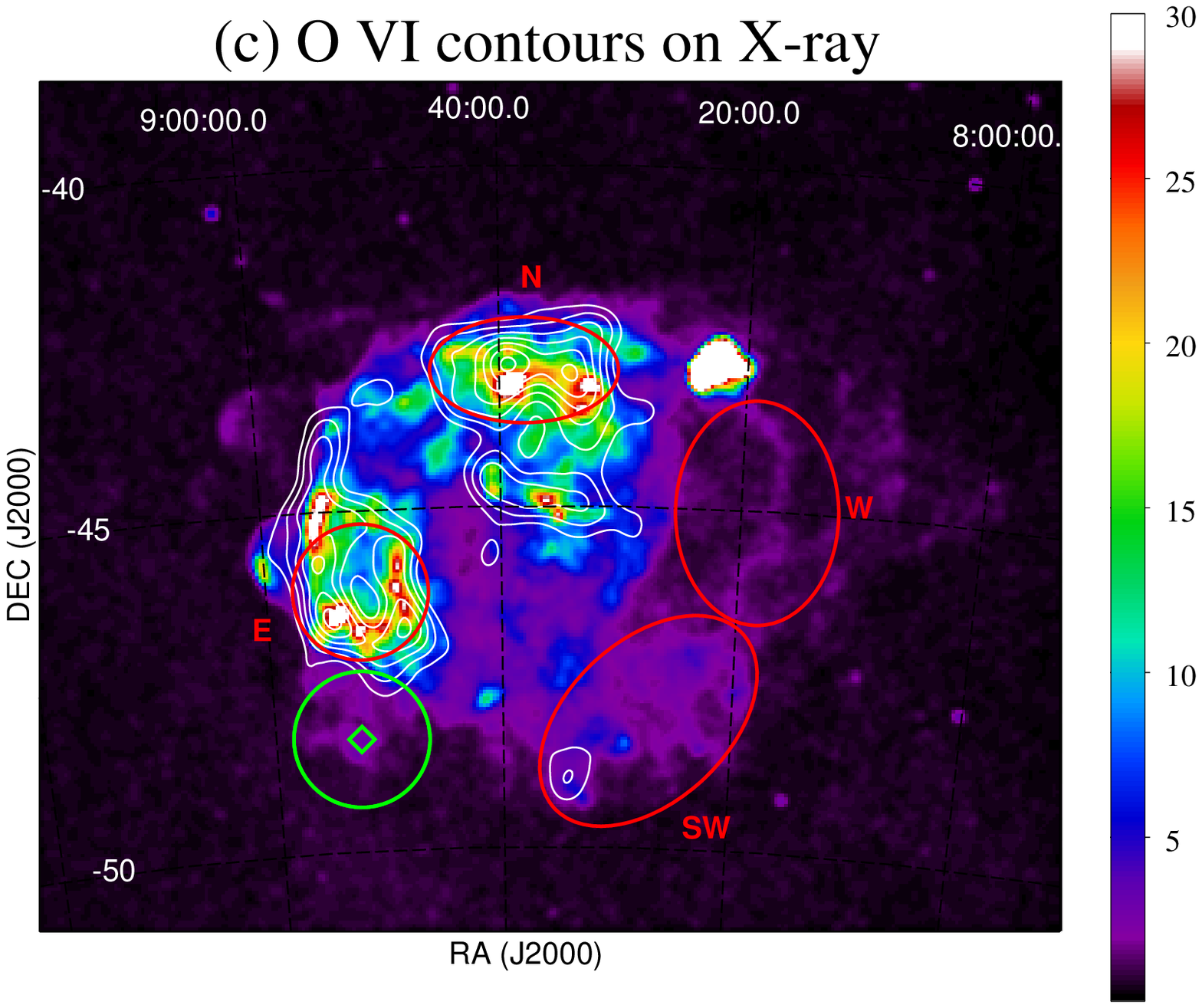}\quad\includegraphics[scale=0.25]{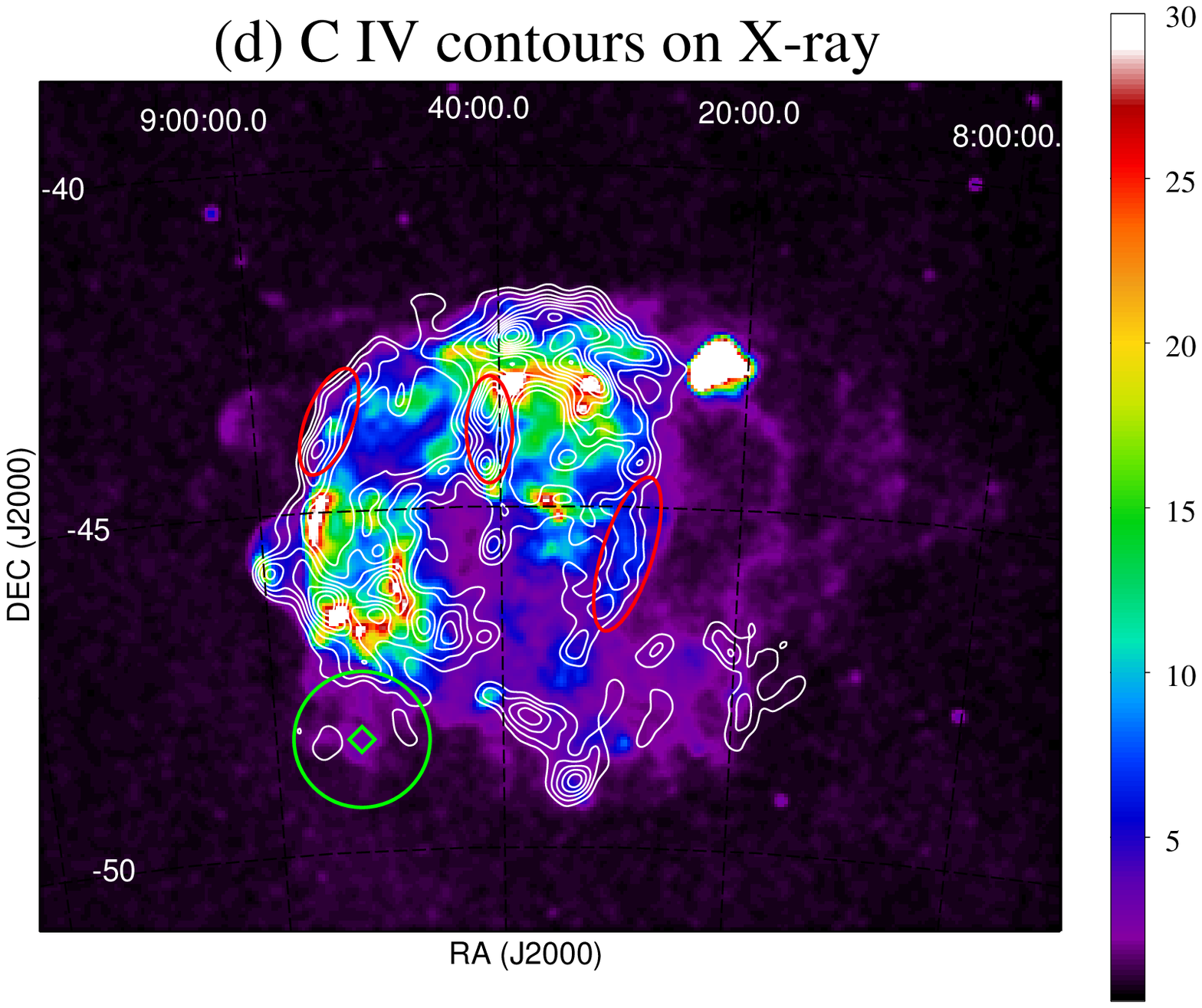}
\par\end{centering}
\textbf{Figure 2.} (a) X-ray \citep{Sushch11} and (b) H$\alpha$ images of the \object{Vela} SNR region. SPEAR/FIMS (c) \ion{O}{6} and (d) \ion{C}{4} contours overlaid on the X-ray (0.1--2.4 keV) images. The H$\alpha$ image in (b) and the X-ray images in (c)--(d) are produced by using the archival data from the SkyView virtual observatory \citep{McGlynn98}. The units of the color bars are (b) rayleigh (10$^{6}/4\pi$ photons s$^{-1}$ cm$^{-2}$ sr$^{-1}$) and (c)--(d) count. The H$\alpha$ contour levels in (b) are from 150 to 300 rayleighs with 30 rayleigh intervals. The SPEAR/FIMS contours in (c) and (d) are same as those in Figures 1(c) and (a), respectively. The diamond symbols denote the position of HD 76161 and circles indicate an H$\alpha$ ring feature surrounding the star in (b)--(d). The four subregions are marked in (c) and three small regions on the filamentary FUV features are so in (d).
\end{figure}

In Figure 2, the FUV images are compared with the {\it ROSAT} All-Sky Survey X-ray and Southern H-Alpha Sky Survey (SHASSA) H$\alpha$ \citep{Gaustad01} images. The archival data of X-ray and H$\alpha$ were adapted from the SkyView virtual observatory \citep{McGlynn98}. As mentioned in Section 1, \citet{Sushch11} suggested that the observed northeast-southwest asymmetry of the \object{Vela} SNR could be due to the $\gamma^{2}$ Velorum SWB, which is clearly shown in Figure 2(a) adapted from them. The X-ray shell is composed of the bright northeast section (hereafter, section XNE) and the dim southwest section (section XSW), which seem to be separated by the boundary of the $\gamma^{2}$ Velorum SWB indicated in the figure. The H$\alpha$ morphology in Figure 2(b) exhibits complex features, none of which show similarity to the X-ray extent of the remnant. This is because the \object{Vela} SNR region contains many objects whose H$\alpha$ intensities overwhelm that of the \object{Vela} SNR. A few circular features seen in the upper-left area of the H$\alpha$ map are the well-known \ion{H}{2} regions, and the large complex features in the lower-right area are parts of the \object{Gum nebula}, known as a very large sphere of ionized gas. In addition, we identified a ring-like H$\alpha$ feature marked with a circle with a radius of 1$\arcdeg$ in Figures 2(b)--(d). At the same position, a faint extended X-ray feature, similarly sized, can be perceived in Figure 2(a), which will be discussed later. In Figures 2(c) and (d), the \ion{O}{6} and \ion{C}{4} contours are overplotted on the {\it ROSAT} X-ray (0.1--2.4 keV) images. In Figure 2(c), the \ion{O}{6} contours are quite similar to the X-ray enhanced regions. This indicates that the two enhanced FUV regions (the north and east enhanced FUV regions) are also the brightest regions in the soft X-ray wavelength domain. However, their peak locations are not exactly matched. In Figure 2(d), most of the \ion{C}{4} contours denote features in section XNE, except the southwest FUV features. FUV filaments A and C seem to form the boundary of this section. The southwest FUV features also appear to envelop another faint extended X-ray region in section XSW.

We designate four subregions in Figure 2(c) for spectral analysis: the north enhanced FUV region (N), the east enhanced FUV region (E), the southwest FUV features (SW), and the west faint region (W). Subregion E was selected to be coincide with the outline of the Vela Jr. SNR, which will be discussed later. Figure 3 shows the SPEAR/FIMS FUV spectra for the whole region and four subregions. The detector background of 0.006 counts s$^{-1}$ \AA$^{-1}$ was subtracted from the spectra \citep{Seon11}. Since the pixels related to bright stars were masked out, most of the FUV continuum is scattered light, by the interstellar dust, of the FUV stellar radiation \citep{Seon11}. The emission-line features are most clearly ascertained in the spectra for subregion N, the most prominent region of the \object{Vela} SNR. The S-channel spectrum confirms the \ion{C}{3} $\lambda$977, \ion{N}{3} $\lambda$991, and \ion{O}{6} $\lambda\lambda$1032, 1038 lines, along with the hydrogen Lyman series ($\gamma$ 973 \AA{}, $\beta$ 1026 \AA{}) originating from the geocoronal airglow. In the L-channel spectrum, the \ion{Si}{4} $\lambda$1394, \ion{O}{4}{]} $\lambda$1404, \ion{N}{4}{]} $\lambda$1486, \ion{C}{4} $\lambda\lambda$1548, 1551 (unresolved), \ion{He}{2} $\lambda$1640.5, and \ion{O}{3}{]} $\lambda\lambda$1661, 1666 (unresolved) lines can be identified. All the lines here have already been detected in \citet{Nishikida06}. For each subregion, the spectra have been fitted to obtain emission-line intensities using the same method which produced the above emission-line images. Assuming $E(\bv)$ = 0.1 \citep{Raymond97,Sankrit03,Wallerstein90} and using the extinction curve of \citet{Cardelli89} with $R_V$ = 3.1, the line intensities were reddening-corrected. The extinction curve of \citet{Cardelli89} differs from that used in \citet{Nishikida06}, but the difference of the reddening correction is $<$10\% for the S-channel and $<$5\% for the L-channel. We extrapolated the extinction curve of \citet{Cardelli89}, which covers only wavelengths $>$1000 \AA, down to the \ion{C}{3} $\lambda$977 and \ion{N}{3} $\lambda$991 lines. However, we note that the currently available extinction curves are highly uncertain in this short wavelength domain. The line luminosities were calculated, adopting a distance of 290 pc to the \object{Vela} SNR \citep{Caraveo01,Dodson03}. Only the lines with a significance of $>$2$\sigma$ are shown in Table 1. The FUV line luminosities for the \object{Cygnus Loop} \citep{Seon06}, along with the X-ray luminosity \citep{Lu00}, are also given for comparison. The \ion{N}{3} $\lambda$991 and \ion{N}{4}{]} $\lambda$1486 luminosities of the \object{Vela} SNR were newly obtained, though their images could not be produced because of their low significance. The \ion{Si}{4} $\lambda$1394 and \ion{O}{4}{]} quintet lines were estimated separately, but the former line was detected at $>$2$\sigma$ significance only in subregions N and E.

To inspect the differences of shock conditions between the subregions, we examined the diagnostic diagrams of line ratios that are sensitive to nonradiative shocks. Figure 4 presents the line ratio diagrams of \ion{Si}{4}/\ion{O}{6} versus \ion{O}{3}{]}/\ion{O}{6} (the left panel) and \ion{Si}{4}/\ion{C}{4} versus \ion{O}{3}{]}/\ion{C}{4} (the right panel), which shows striking differences between subregions N and E. In the figure, only the pixels with the line ratios of $>$2$\sigma$ significance, are plotted. All the pixels for subregion E clearly concentrate at the low-ratio ranges compared with those for subregion N. Especially, the \ion{O}{3}{]}/\ion{O}{6} values for subregion E (roughly 0.05--0.2) are about twice lower than those for subregion N (roughly 0.1--0.5). We will further discuss the implications of these differences later with regards to the contribution of nonradiative shocks in subregion E.

\begin{figure}
\begin{centering}
\includegraphics[scale=0.4]{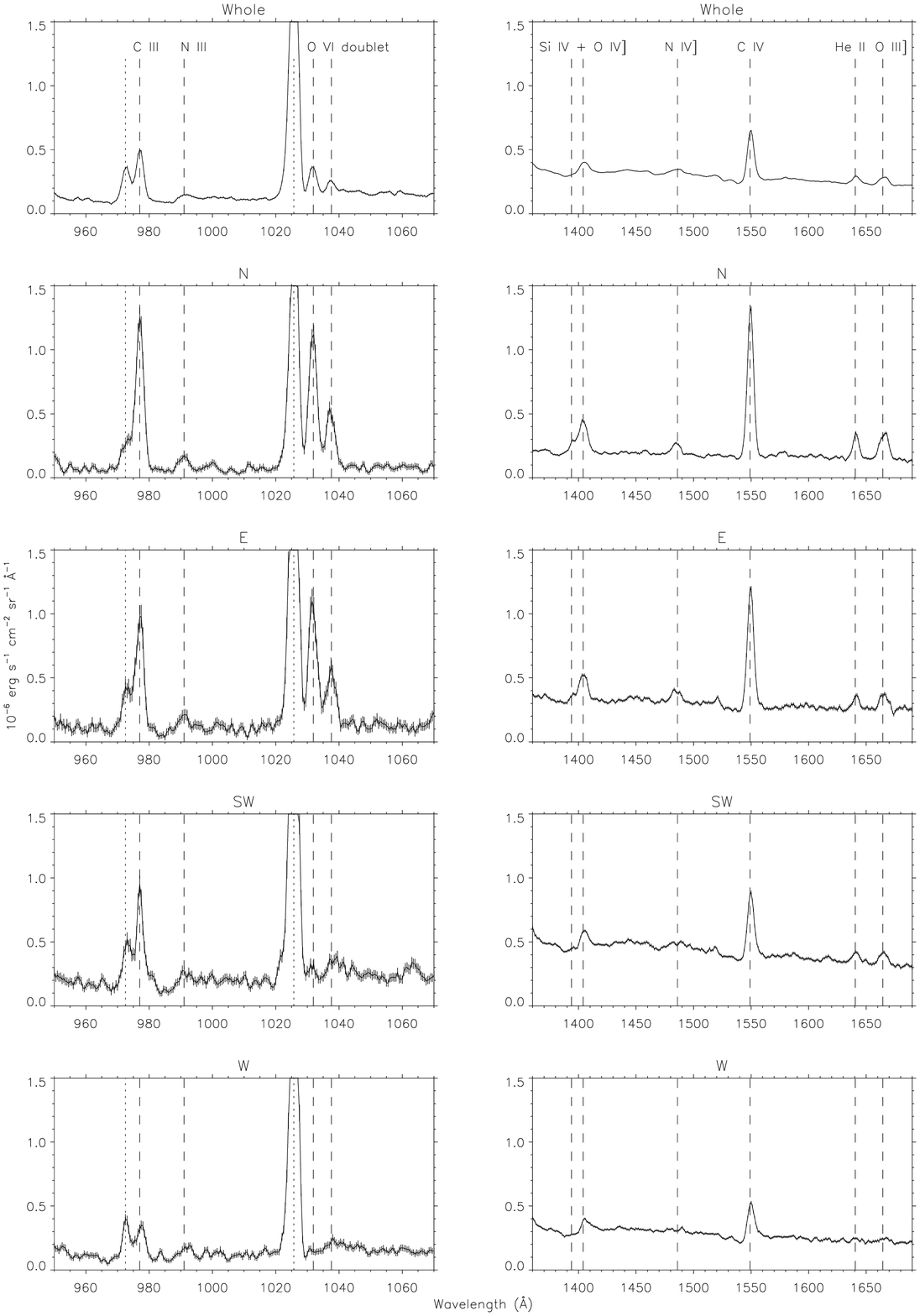}
\par\end{centering}
\textbf{Figure 3.} SPEAR/FIMS S-channel ($\it left$) and L-channel ($\it right$) spectra (with 1 $\sigma$ error bars) from the whole region and four subregions indicated in Figure 2(c). The spectra are binned with 0.5 (S-channel) and 1 (L-channel) \AA\ intervals and smoothed with a boxcar average of 3 bins. The positions of the identified emission lines are given by dashed lines. The hydrogen Lyman series ($\gamma$ 973 \AA, $\beta$ 1026 \AA) originating from the geocorona are also marked by dotted lines.
\end{figure}

\begin{deluxetable*}{lcccccc}
\tabletypesize{\footnotesize}
\tablewidth{0pt}
\tablecaption{FUV Line Luminosities and Comparisons with the Cygnus Loop\label{tbl-1}}
\tablehead{
\colhead{Species} & \colhead{N} & \colhead{E} & \colhead{SW} & \colhead{W} & \colhead{Whole} & \colhead{Cygnus Loop}
}
\startdata
\ion{C}{3} $\lambda$977 & 2.07 $\pm$ 0.10 & 1.41 $\pm$ 0.10 & 2.19 $\pm$ 0.16 & 0.69 $\pm$ 0.08 & 21.84 $\pm$ 0.49 & 8.82\tablenotemark{b} \\
\ion{N}{3} $\lambda$991 & 0.23 $\pm$ 0.04 & 0.27 $\pm$ 0.05 & 0.42 $\pm$ 0.10 & 0.23 $\pm$ 0.06 & 3.01 $\pm$ 0.33 & \nodata \\
\ion{O}{6} $\lambda\lambda$1032, 1038 & 2.12 $\pm$ 0.09 & 1.79 $\pm$ 0.10 & 0.42 $\pm$ 0.12 & 0.20 $\pm$ 0.07 & 14.81 $\pm$ 0.45 & 15.0\tablenotemark{b} \\
\ion{Si}{4} $\lambda\lambda$1394, 1403 & 0.16 $\pm$ 0.02 & 0.07 $\pm$ 0.02 & \nodata & \nodata & \nodata & \multirow{2}{*}{0.66 $\pm$ 0.06}\tablenotemark{b} \\
\ion{O}{4}{]} $\lambda$1404 & 0.57 $\pm$ 0.03 & 0.41 $\pm$ 0.03 & 0.44 $\pm$ 0.05 & 0.28 $\pm$ 0.03 & 6.13 $\pm$ 0.17 & \\
\ion{N}{4}{]} $\lambda$1486 & 0.11 $\pm$ 0.01 & 0.09 $\pm$ 0.01 & 0.06 $\pm$ 0.02 & \nodata & 1.47 $\pm$ 0.07 & \nodata \\
\ion{C}{4} $\lambda\lambda$1548, 1551 & 1.88 $\pm$ 0.02 & 1.38 $\pm$ 0.02 & 1.56 $\pm$ 0.03 & 0.76 $\pm$ 0.02 & 20.28 $\pm$ 0.12 & 4.47 $\pm$ 0.14\tablenotemark{b} \\
\ion{He}{2} $\lambda$1640.5 & 0.28 $\pm$ 0.01 & 0.14 $\pm$ 0.02 & 0.23 $\pm$ 0.03 & 0.07 $\pm$ 0.02 & 2.75 $\pm$ 0.10 & 0.68 $\pm$ 0.06\tablenotemark{b} \\
\ion{O}{3}{]} $\lambda\lambda$1661, 1666 & 0.39 $\pm$ 0.02 & 0.20 $\pm$ 0.02 & 0.33 $\pm$ 0.03 & 0.10 $\pm$ 0.02 & 3.56 $\pm$ 0.12 & 0.65 $\pm$ 0.08\tablenotemark{b} \\
X-ray & \nodata & \nodata & \nodata & \nodata & 3.0\tablenotemark{a} & 3.59\tablenotemark{b} \\
\enddata
\tablecomments{All line luminosities are in units of 10$^{35}$ erg s$^{-1}$. The first four subregions are indicated in Figure 2(c). The values for the \object{Vela} are calculated adopting a distance of 290 pc \citep{Caraveo01,Dodson03} and reddening-corrected assuming $E(\bv)$ = 0.1 \citep{Raymond97,Sankrit03,Wallerstein90}. Only line luminosities determined with $>$2$\sigma$ significance are listed.}
\tablenotetext{a}{The value is from \citet{Lu00} and scaled to a distance of 290 pc. The corresponding energy band is 0.1--2.5 keV.}
\tablenotetext{b}{The values are from \citet{Seon06}. The corresponding energy band for the X-ray value is 0.1--4.0 keV.}
\end{deluxetable*}

The velocity for an individual shock could not be estimated because of the limited spatial resolution of the SPEAR/FIMS data. We, therefore, attempted to model the mean FUV line ratios obtained from three small regions selected on FUV filaments A, B, and C, as marked in Figure 2(d), with the values expected from mixture of shocks with a power-law velocity distribution \citep{Vancura92}. The FUV line ratios expected from each shock velocity were calculated using the MAPPINGS III models \citep{Allen08}. We used only the \ion{C}{3}, \ion{O}{6}, \ion{C}{4}, \ion{He}{2} and \ion{O}{3}{]} to \ion{O}{4}{]} ratios because the \ion{N}{3}, \ion{Si}{4}, and \ion{N}{4}{]} lines were not detected strongly enough for reliable estimation. For various values of shock velocity range, preshock density, preshock magnetic field, and abundance, we tried to determine the power-law indexes of velocity distribution which predict the observed line ratios. However, for any of the three regions, even the best fit results did not match the data well. In fact, in estimating the velocities of radiative shocks, we need to take into account the effects of X-ray emitting gas, nonradiative shocks, and resonance scattering \citep{Raymond97}. Although the \ion{O}{6} line is the most sensitive to the velocities of radiative shocks, very close morphological resemblance between the \ion{O}{6} and soft X-ray images, as can be seen in Figure 2(c), seems to indicate a contribution to the \ion{O}{6} intensity from the X-ray emitting gas. The \ion{He}{2} line intensity can be contributed to by nonradiative shocks \citep{Raymond83,Long92,Hester94}. Resonance scattering affects the intensities of resonance lines such as \ion{O}{6}, \ion{C}{3} and \ion{C}{4}  \citep{Raymond81,Raymond97}. Moreover, the \ion{C}{3} $\lambda$977 line has relatively large uncertainties in the reddening-correction. Eventually, only the \ion{O}{3}{]} to \ion{O}{4}{]} ratio, which is also independent of elemental abundances, remains free from the above effects. However, this ratio is not very sensitive to the shock velocities. When the \ion{O}{3}{]} to \ion{O}{4}{]} ratios are used, we obtained the lowest power-law indexes for the region on FUV filament C, implying relatively larger contribution of slow shocks than the regions on FUV filaments A and B.

\section{DISCUSSION}

We presented the new FUV emission-line images of the \object{Vela} SNR with higher resolutions than those in \citet{Nishikida06} by using a new SPEAR/FIMS data set, which included more orbits and were processed with better attitude knowledge than \citet{Nishikida06}. They revealed not only detailed bright FUV features, but also faint FUV features that were not observed in \citet{Nishikida06}, such as FUV filament C and the FUV features in subregion SW. In particular, the new \ion{C}{3} and \ion{O}{6} images revealed new FUV features in the regions uncovered in \citet{Nishikida06}, such as the FUV features in subregions E and SW. Compared with Figure 2 of \citet{Nishikida06}, the new images are found to agree well with the old images for the regions presented in \citet{Nishikida06}.

We now compare the FUV line luminosities with those calculated in \citet{Nishikida06} by scaling their results with the same reddening curve and distance adopted in the present study. The \ion{C}{4}, \ion{He}{2}, and \ion{O}{3}{]} luminosities for the whole region in Table 1, are found to be $\sim$60--85\% higher than those in \citet{Nishikida06}. \citet{Nishikida06} assumed a diameter of 8$\arcdeg$, but we obtained the luminosities from the whole area displayed in Figure 1. The discrepancy implies that the \object{Vela} SNR has a significant amount of FUV emissions outside the 8$\arcdeg$ circular area. On the other hand, the \ion{C}{3} and \ion{O}{6} luminosities for the whole region are $\sim$45--55\% lower than those in \citet{Nishikida06}. Moreover, the \ion{O}{6}/\ion{C}{3} ratio is significantly different from their result. \citet{Nishikida06} applied the S-channel line intensities obtained from the brightest part of the remnant to the entire SNR, which gave rise to an overestimation of the S-channel line luminosities. Moreover, the \ion{O}{6} emissions are more concentrated in the region covered in \citet{Nishikida06}, than the \ion{C}{3} emissions, as can be seen in Figures 1(b) and (c). Therefore, the \ion{C}{3} luminosity was less overestimated than the \ion{O}{6}, yielding an overestimated \ion{O}{6}/\ion{C}{3} ratio of $\sim$1 in \citet{Nishikida06}. The present result of \ion{O}{6}/\ion{C}{3} = 0.68 is much closer to the value (0.6) reported by \citet{Blair95}.

\begin{figure}[t]
\begin{centering}
\includegraphics[scale=0.2]{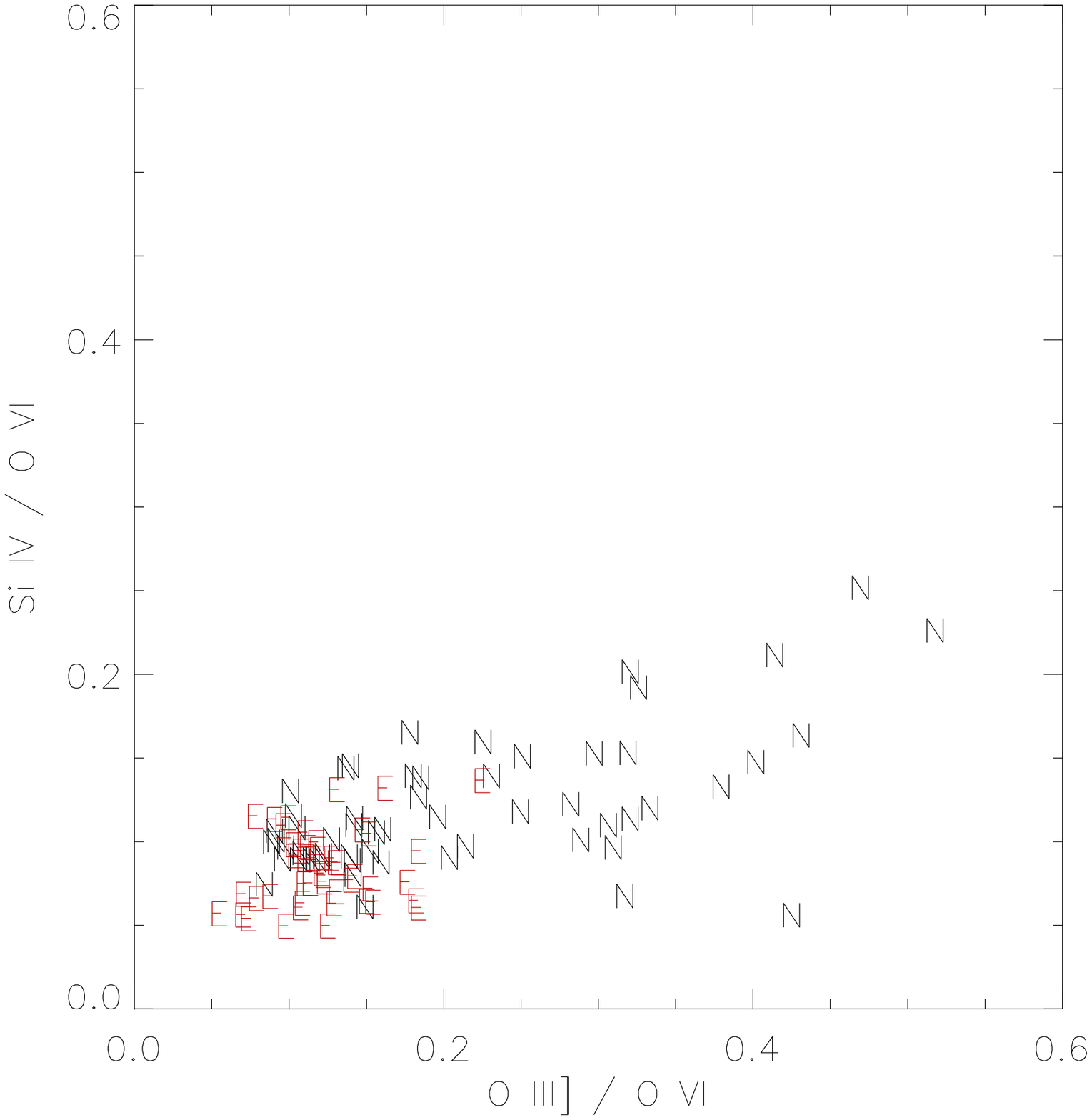}\quad\includegraphics[scale=0.2]{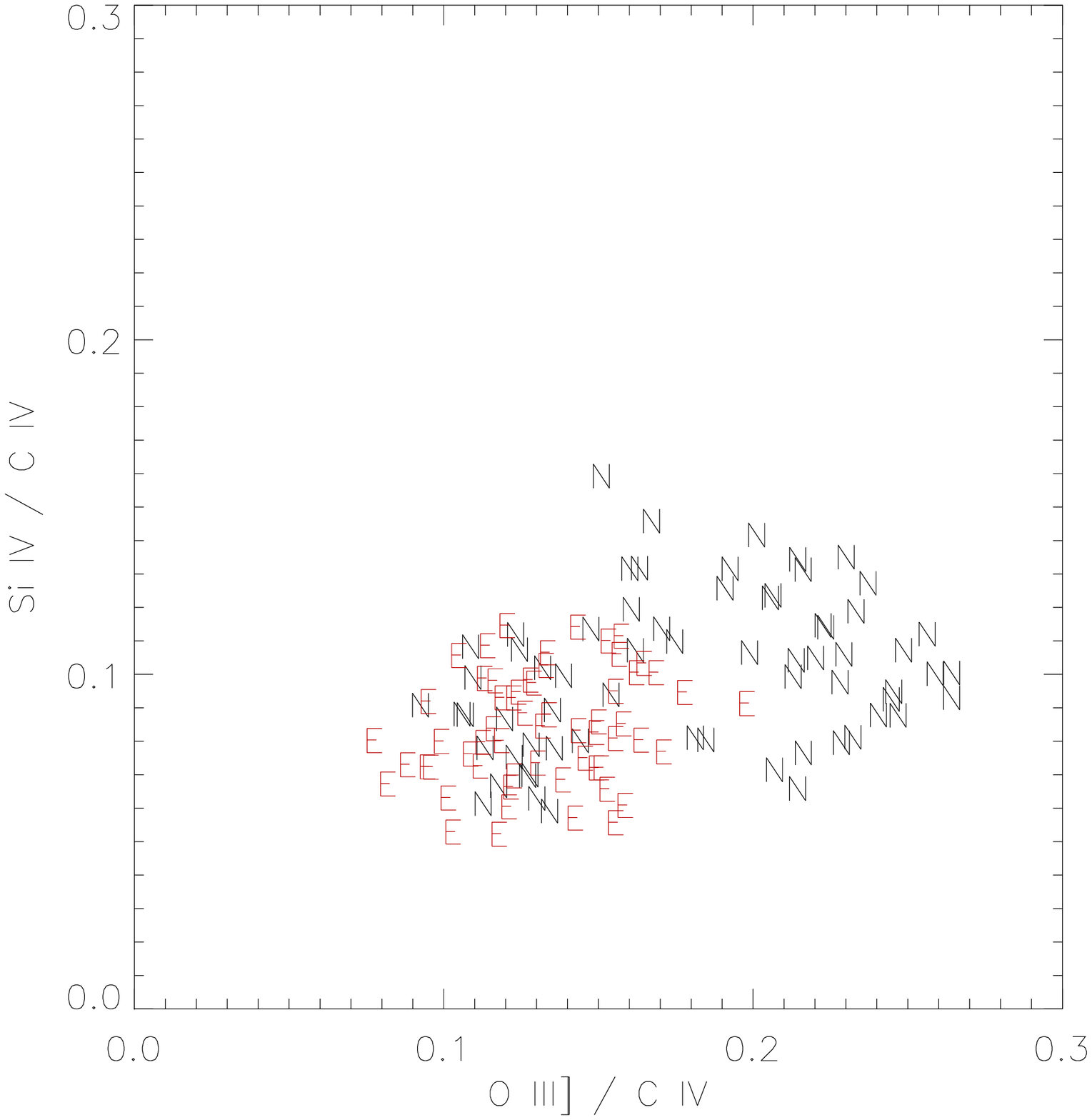}
\par\end{centering}
\textbf{Figure 4.} Diagnostic line ratio diagrams for subregions N and E. The data points are marked by the name of subregion where each data point lies (To show distinctly, all points for subregion E are colored red in the online version).
\end{figure}

It was noted in \citet{Nishikida06} that the enhanced FUV feature in subregion N appears at all FUV emission-line images and the soft X-ray image, although the peak positions are not exactly coincidental. The mismatch is more clearly shown in Figures 2(c) and (d). We also note that the peak locations of the \ion{O}{6} and \ion{C}{4} images are slightly different from each other. The peak of \ion{C}{4} appears at the outer region of that of \ion{O}{6}, which is located outside of the X-ray peak. All the other FUV emission-line images also show slight shifts in their peaks, although their overlaid images have not been displayed. \citet{Bocchino00} found stratification of the X-ray, H$\alpha$, and [\ion{O}{3}{]} $\lambda$5007 emissions in a small part of subregion N. They concluded that this is direct evidence that the shock is propagating along a medium with increasing density gradient. This is supported by the results of the \ion{H}{1} \citep{Dubner98} and CO \citep{Moriguchi01} studies, showing that the \object{Vela} SNR is interacting with dense clouds in this region. Moreover, the CO morphology \citep{Moriguchi01} shows counterparts of the two concentric arc-like features, which are also seen in the \ion{C}{4} image. They suggested that the outer arc-like feature can be an expanding shock front viewed edge-on. Because of the limited spatial resolution, only the \ion{O}{4}{]} image but the \ion{C}{4} exhibits a marginal sign of the outer arc-like feature. Subregion E is also one of the brightest regions in all the FUV emission-line and X-ray images, although its detailed structures somewhat differ in different wavelengths. The peak patterns in the \ion{O}{6} and soft X-ray images resemble a ring-like feature, as can be seen in Figure 2(c). The southern circular part of the ring feature coincides with the outline of the Vela Jr. SNR, known as another younger SNR \citep{Aschenbach98}. Because of the superposition of the strong emissions from the \object{Vela} SNR, only the nonthermal gamma-ray \citep{Aharonian07,Katagiri05}, hard X-ray \citep{Aschenbach98,Slane01}, and radio \citep{Combi99,Duncan00} emissions have been confirmed to be associated with the Vela Jr. SNR so far. We also note that no \ion{H}{1} \citep{Dubner98}, CO \citep{Moriguchi01},
or dust \citep{Nichols04} counterparts for subregion E have been found. Therefore, it would be important to examine the possible association of the ring feature with the Vela Jr. SNR, although the morphological similarity at the southern part of the ring feature does not necessarily indicate physical association of the ring feature with the Vela Jr. SNR.

The contribution of the Vela Jr. SNR to the FUV emissions observed in subregion E may be investigated by noting that nonradiative shocks are more dominant in younger SNRs. The Vela Jr. SNR is known to be young ($\lesssim1500$ yr; Aschenbach 1998, 1700--4300 yr; Katsuda et al. 2008) so that most shocks of the remnant are expected to be nonradiative, in contrast to the case of the \object{Vela} SNR where radiative shocks are dominant. The FUV spectrum of a nonradiative shock in the \object{Cygnus Loop} has been obtained by \citet{Long92}. The \ion{O}{6} $\lambda\lambda$1032, 1038, \ion{C}{4} $\lambda\lambda$1548, 1551, and \ion{He}{2} $\lambda$1640.5 emission lines were detected strongly, but the \ion{Si}{4} $\lambda\lambda$1394, 1403 and \ion{O}{3}{]} $\lambda\lambda$1661, 1666 lines were very weak, as can be seen in Table 2 of \citet{Long92}. The \ion{C}{3} $\lambda$977 and \ion{O}{4}{]} $\lambda$1404 lines were observed to have intermediate intensities. This property of nonradiative shocks accords well with the distinct differences in line ratios between subregions N and E, as shown in Figure 4. The ratio between the weakest and the strongest lines (\ion{O}{3}{]} and \ion{O}{6}, respectively) in nonradiative shocks, as observed in the \object{Cygnus Loop}, shows a clear distinction between subregions N and E. The ratios including the secondly weakest and strongest lines (\ion{Si}{4} and \ion{C}{4}, respectively) in nonradiative shocks also show differences in Figure 4. The relatively low line ratios in subregion E may be attributed to the contribution of nonradiative shocks from the Vela Jr. SNR. Assuming that the emission lines from radiative shocks of the \object{Vela} SNR are not significantly different between subregions N and E, we could estimate the contribution of nonradiative shocks from the Vela Jr. SNR. The fact that the maximum value of  the \ion{O}{3}{]}/\ion{O}{6} ratio for subregion E is only about half of that for subregion N in Figure 4 suggests that the contribution from the Vela Jr. SNR to the spectrum of subregion E can be up to 50\%. We can then constrain the distance of the Vela Jr. SNR  because of strong dust extinction at FUV wavelengths hampering the detection of FUV emissions from a distant object at a distance of $>$1 kpc near the Galactic plane. Several studies \citep{Aschenbach98,Bamba05,Katsuda08} estimated the distance of the Vela Jr. SNR to be less than 1 kpc, whereas \citet{Slane01} suggested that the Vela Jr. SNR might be located near the Vela Molecular Ridge at a distance of 1--2 kpc. Our FUV results support the former estimations of the distance to the Vela Jr. SNR.

FUV filament A detected in all the FUV emission-line images exactly delineates the northeast border of the X-ray emission, as can be seen in Figure 2(d). The \ion{H}{1} \citep{Dubner98} and CO \citep{Moriguchi01} studies showed that dense clouds envelop this sharp boundary of the X-ray image, suggesting a strong interaction with the \object{Vela} SNR. Based on less prominent optical filaments in this region than in the north region, they proposed that the clouds might be located on the near side of the remnant and contain dust responsible for increased extinction. However, if this is the case, FUV filament A would be too severely obscured to be observable. Hence, the relative weakness of the optical emissions at the northeast border is most likely due to other factors such as high shock velocities. FUV filament B is the most prominent FUV filament, although the \ion{He}{2} and \ion{O}{3}{]} lines are brighter along FUV filament C. The soft X-ray (also the \ion{O}{6} line) emissions appear as a bright knot at the southernmost part of FUV filament B. Previous FUV studies \citep{Raymond81,Raymond97,Sankrit01} detected various FUV emission lines from some small regions near the bright X-ray knot and concluded that the FUV emissions come from $\sim$132--180 km s$^{-1}$ shocks. FUV filament C coincides well with the sharp boundary of the northeast-southwest asymmetry of the X-ray shell, as can be seen in Figures 2(a) and (d). This result confirms the existence of asymmetry and is compatible with the suggestion of \citet{Sushch11} that the asymmetry could be due to the $\gamma^{2}$ Velorum SWB, whose shell must contain denser material. Indeed, FUV filament C, out of the three FUV filaments, is brightest in the lower ionization emission lines such as the [\ion{O}{2}{]} $\lambda$3727 \citep{Miller73} and [\ion{O}{3}{]} $\lambda$5007 \citep{Nichols04}, implying the predominance of shocks with lower velocities. This is also compatible with our results of the \ion{O}{3}{]}/\ion{O}{4}{]} ratios (although this ratio is not sensitive to the shock velocities) on the three FUV filaments, confirming the dominance of slower shocks in FUV filament C than in FUV filaments A and B.

Apart from the above FUV features, the FUV features in subregion SW lie in the dim southwest section of the asymmetrical X-ray shell (section XSW). \citet{Sushch11} argued that section XSW is dimmer and more extended than section XNE because of more rarefied medium inside the $\gamma^{2}$ Velorum SWB. Nevertheless, the brightest part of subregion SW, seen in all the FUV emission-line images, has radio continuum \citep{Bock98} and CO \citep{Moriguchi01} counterparts. A detailed comparison shows that a faint filamentary feature of the radio continuum (843 MHz) and arc-like CO feature surround the west and the east sides of the FUV peak, respectively. Furthermore, the faint west part of subregion SW, which appears clearly in the \ion{C}{3}, \ion{O}{4}{]}, \ion{C}{4}, \ion{He}{2}, and \ion{O}{3}{]} images, seems to surround a faint extended X-ray region, as can be seen in Figure 2(d). Comparing the spectra for subregions SW and W in Figure 3 shows that the \ion{C}{3}, \ion{O}{4}{]}, \ion{C}{4}, \ion{He}{2}, and \ion{O}{3}{]} lines are stronger in subregion SW than subregion W. The average intensities of the \ion{C}{3},
\ion{O}{4}{]}, \ion{C}{4}, \ion{He}{2}, and \ion{O}{3}{]} lines in subregion SW are $\sim$2.8, $\sim$1.4, $\sim$1.8, $\sim$3.1, and $\sim$3.0 times higher than those in subregion W, respectively. Therefore the morphology and brightness of subregion SW indicate that this region may have slightly denser preshock medium within the dim section XSW.

\citet{Lu00} noticed some isolated high temperature X-ray features in their mean temperature map of the \object{Vela} SNR (see Figure 11 in their paper). They numbered those features from (1) to (5) and then discussed each feature one by one. Feature (1) was related to Shrapnel D, and features (2) and (3) were related to the Vela Jr. SNR. They also mentioned that feature (4) lying outside the shock wave boundary of the remnant might be an artifact by high background, and feature (5) inside the boundary may be a result of the interaction of a shock wave with some low density medium. We note that features (4) and (5) coincide well with the extended X-ray feature associated with the H$\alpha$ ring in Figure 2. Furthermore, two local peaks of \ion{C}{4} emission appear at the positions of features (4) and (5), as can be seen in Figure 2(d), which absolutely points out that feature (4) must not be an artifact. Considering the whole extent of the H$\alpha$ ring, feature (3) might be also related to the H$\alpha$ ring and its extended X-ray counterpart.

The H$\alpha$ ring would be in direct contact with the Vela SNR. This can be compared with the Gemini H$\alpha$ ring, which is also in physical contact with the Monogem ring SNR \citep{Kim07}. The X-ray morphology of the Monogem ring is strongly distorted inward and the \ion{C}{4} emission peaks along the border between the X-ray remnant and Gemini H$\alpha$ ring. However, the X-ray and \ion{C}{4} emission features in the present H$\alpha$ ring are rather extended compared to those of the Gemini H$\alpha$ ring. If the contact plane between the H$\alpha$ ring and remnant is viewed edge-on, the morphologies of the X-ray and \ion{C}{4} emissions would appear as found in the Gemini H$\alpha$ ring. On the other hand, if our line of sight departs from the edge-on case, the extended X-ray and \ion{C}{4} emissions, as in the present H$\alpha$ ring in Figure 2, would be observed. Therefore, we conclude that the H$\alpha$ ring is in contact with the near or far side of the \object{Vela} SNR.

The H$\alpha$ ring may be due to an \ion{H}{2} region and/or SWB formed by a nearby early-type star. If the H$\alpha$ ring is due to an \ion{H}{2} region, we can estimate the spectral type of the central ionizing star of the \ion{H}{2}
region and ambient density from the observed H$\alpha$ intensity and size of the ring. The H$\alpha$ intensity of the H$\alpha$ ring is calculated to be 85 rayleighs by subtracting the background value (100 rayleighs). The background region was selected to be an adjacent region with the same size as the H$\alpha$ ring's. The radius (1$\arcdeg$) of the H$\alpha$ ring is $\sim$5 pc at the distance to the \object{Vela} SNR of 290 pc. Adopting the case B relation between the H$\alpha$ luminosity ($L_{\mathrm H\alpha}$) and the Hydrogen-ionizing (Lyman continuum) photon luminosity ($L_{\rm Lyc}$), i.e., $L_{\mathrm H\alpha}$ = 0.36$L_{\rm Lyc}$ \citep{Martin88, Wood99}, we obtain  $\log L_{\rm Lyc}$ (photons s$^{-1}$) = 47.23  for the H$\alpha$ intensity  and radius of the ring. \citet{Seon07} provides $L_{\rm Lyc}$ and the Str\"{o}mgren radii of \ion{H}{2} regions expected for various stellar types, which are calculated using the effective temperatures, gravities, and stellar radii from stellar evolutionary models (Strai$\check{\rm z}$ys \& Kuriliene 1981) and adopting Kurucz stellar atmosphere model \citep{Castelli03}. From these calculations, the source of the H$\alpha$ ring is expected to be a B0 or B1-type star in a medium with a density of $\sim$6.5 cm$^{-3}$. \citet{Sushch11} estimated the ISM density in the direction of the \object{Vela} SNR based on the dynamics of expansion of the \object{Vela} SNR, $\gamma^{2}$ Velorum SWB, and \object{Gum nebula}. The resulting values differ widely: $\leq$ 0.01 cm$^{-3}$, $\sim$20 cm$^{-3}$, and 0.07 cm$^{-3}$, respectively, indicating highly inhomogeneous media of the \object{Vela} SNR region. We note that there exists a bright star near the center of the H$\alpha$ ring: HD 76161 denoted by a diamond symbol in Figures 2(b)--(d). {\it Hipparcos} distance of HD 76161 is 333 $\pm$ 58 pc \citep{ESA97} and its spectroscopic distance determined by \citet{Nichols04} is 296 pc, which is similar to the distance to the \object{Vela} SNR of 290 pc \citep{Caraveo01,Dodson03}. This star is the only bright point source detected within the H$\alpha$ ring in the SPEAR/FIMS S-channel (900--1150 \AA{}) continuum map. However, the spectral type of HD 76161 is too late (B3V-type; ESA 1997) to be the ionizing source of the H$\alpha$ ring. The B3V-type star can produce an \ion{H}{2} region with the H$\alpha$ intensity of 0.04 rayleighs, which is far less ($\sim$0.05\%) than the observed value (85 rayleighs). As discussed above, the H$\alpha$ ring may be in direct contact with the \object{Vela} SNR and the $\gamma^{2}$ Velorum SWB, which would enhance the H$\alpha$ intensity of the ring. However, we may need further studies for a definite conclusion about the origin of the ring.

The FUV line luminosities, such as the \ion{C}{3}, \ion{O}{6}, \ion{O}{4}{]}, \ion{C}{4}, or \ion{O}{3}{], calculated over the whole remnant in Table 1, exceed the 0.1--2.5 keV X-ray luminosity by a factor of 1.2--7.3. As discussed in \citet{Nishikida06}, these results ascertain the importance of the FUV emission lines as an efficient cooling channel for the \object{Vela} SNR. Based on the lower \ion{O}{6}/\ion{C}{3} ratio of the \object{Vela} SNR, \citet{Blair95} and \citet{Nishikida06} concluded that the \object{Vela} SNR has a relatively large fraction of slower shocks compared with the \object{Cygnus Loop}. This conclusion is corroborated by the comparison in Table 1. Indeed, while the highest ionization FUV line (\ion{O}{6}) and X-ray luminosities are similar to those of the \object{Cygnus Loop}, all the other FUV line luminosities are much larger than those of the \object{Cygnus Loop}.

We have checked the SPEAR/FIMS FUV detections for Shrapnels A--F and the Puppis A SNR. Shrapnels A and B were considered to originate from supernova ejecta by X-ray studies \citep{Strom95,Miyata01,Katsuda06,Yamaguchi09}. On the other hand, Shrapnel D was detected even in the optical and FUV wavelength domains, and there are a few different explanations for its origin \citep{Redman00,Redman02,Plucinsky02,Sankrit03,Katsuda05}. Figure 1 shows the morphologies certainly related to Shrapnel D in the \ion{C}{3}, \ion{O}{6}, \ion{O}{4}{]}, \ion{C}{4}, and \ion{O}{3}{]} images. Also, we newly report the FUV detections from Shrapnel E. In Figure 1(a), the \ion{C}{4} emission exhibits a filamentary feature running east-west around $\alpha \sim 8^{\mathrm h}5^{\mathrm m}, \delta \sim -45\arcdeg30\arcmin$ although it is too weak to be included in the contours (nevertheless, it has S/N $>$3). This filamentary feature corresponds well with the X-ray southern boundary of Shrapnel E in Figure 2(a). Moreover, there is an \ion{O}{4}{]} feature at the similar position, as can be seen in Figure 1(e). Although no FUV feature is found near Shrapnels A--C and F in Figure 1, the \ion{C}{4} emission line is clearly detected in the spectra extracted from regions Shrapnel A--C. However, its physical relationship with the Shrapnels is not clear at this stage. Finally, we could not find any evidence on the FUV line emissions from the Puppis A SNR (around $\alpha \sim 8^{\mathrm h}23^{\mathrm m}, \delta \sim -43\arcdeg$) with the rather limited spatial resolution in the present study.

\section{CONCLUSIONS}

The newly processed SPEAR/FIMS data have allowed us to obtain higher-resolution FUV emission-line images of the \object{Vela} SNR compared with \citet{Nishikida06}, particularly the \ion{C}{4} $\lambda\lambda$1548, 1551 image. The incomplete images of the \ion{C}{3} $\lambda$977 and \ion{O}{6} $\lambda\lambda$1032, 1038 lines have also been updated to cover the whole remnant. In addition, the \ion{He}{2} $\lambda$1640.5 and \ion{O}{3}{]} $\lambda\lambda$1661, 1666 images have been newly presented, and the blended \ion{Si}{4}+\ion{O}{4}{]} image has been deblended into separate \ion{Si}{4} $\lambda$1394 and \ion{O}{4}{]} $\lambda$1404 images. The FUV emission-line luminosities of the \object{Vela} SNR have been estimated more accurately than in \citet{Nishikida06}. In particular, the \ion{C}{3} and \ion{O}{6} luminosities, which were overestimated in \citet{Nishikida06}, were updated. The \ion{N}{3} $\lambda$991, \ion{N}{4}{]} $\lambda$1486, and deblended \ion{Si}{4} and \ion{O}{4}{]} luminosities have been newly calculated as well. Diagnostic line ratio diagrams for two enhanced FUV regions have revealed the outstanding differences between them. The line ratios, which are sensitive to nonradiative shocks, for the east enhanced FUV region concentrate at the lowest values. This result can be explained by the contribution of another younger SNR, the Vela Jr. SNR, which is overlapped with the east enhanced FUV region. If this is the case, our study might be the first detection of the FUV emission lines associated with the Vela Jr. SNR, suggesting that the Vela Jr. SNR is not far from us. The comparison of the improved FUV images with X-ray images has revealed a FUV filamentary feature corresponding with the boundary of the northeast-southwest asymmetry of the X-ray shell. The predominance of shocks with relatively low velocities on the FUV filament accords well the previous proposal that the observed asymmetry of the \object{Vela} SNR could be due to the $\gamma^{2}$ Velorum SWB. Within the dim southwest section of the asymmetrical X-ray shell, the southwest FUV features surrounding a faint extended X-ray region have been identified, and this region was found to have higher FUV intensities than the rest of the region of this section. This implies that the \object{Vela} SNR is interacting with slightly denser ambient medium in the region including the southwest FUV features than in the rest of the region of the southwest section. From a comparison with the H$\alpha$ image, we have identified a ring-like H$\alpha$ feature overlapped with an extended hot X-ray feature of similar size and two local peaks of the \ion{C}{4} emission. These multi-wavelength morphologies are expected when the H$\alpha$ ring is in direct contact with the near or far side of the \object{Vela} SNR.

\acknowledgements{}

We thank the referee, Dr. John Raymond, for his valuable comments, which significantly improved the manuscript. SPEAR/FIMS was supported by NASA Grant NAG5-5355 and the Korean Ministry of Science and Technology. I.-J. Kim, K.-I. Seon, and W. Han were supported by a National Research Foundation of Korea grant funded by the Korean government. K. W. Min was supported by Basic Science Research Program and National Space Laboratory Program through the National Research Foundation of Korea (NRF) funded by the Ministry of Education, Science and Technology.

\end{document}